\documentclass[prd,aps,preprint,showpacs,preprintnumbers,nofootinbib]{revtex4}
\usepackage{epsfig,footnote}
\usepackage{ulem}
\usepackage{color}
\usepackage{array}
\usepackage{amssymb}
\usepackage{amsmath}
\usepackage{graphicx,subfigure}
\usepackage{longtable}
\usepackage{verbatim}
\usepackage{amsfonts}
\usepackage{hyperref}
\usepackage{cancel}
\newcommand{\degree}{^\circ} 
\begin{document}
\title{Extraction of the CKM phase $\gamma$ from charmless two-body $B$ meson decays  }
\author{Si-Hong Zhou}
\affiliation{School of Physical Science and Technology,
Inner Mongolia University, Hohhot 010021, China}
\author{Cai-Dian L\"u}
\affiliation{ Institute of High Energy Physics, CAS, Beijing 100049, China }
\affiliation{ School of Physics, University of Chinese Academy of Sciences, Beijing 100049, China
}
\date{\today}
\begin{abstract}
Utilizing all the experimental measured charmless $B \to PP$, $PV$ decay modes, 
where $P(V)$ denotes a light pseudoscalar (vector) meson, we extract the CKM 
angle $\gamma$ by global fit. All the unknown hadronic parameters are fitted 
with $\gamma$ together from experimental data, so as to make the approach 
least model dependent. The different contributions for various decay modes 
are classified by topological weak Feynman diagram amplitudes, which are to 
be determined by the global fit. To improve the precision of  this approach, 
we consider flavor SU(3) breaking effects of topological diagram amplitudes 
among different decay modes by including the form factors and decay constants. 
The fitted result for CKM angle $\gamma$ is $(69.8 \pm 2.1 \pm 0.9) ^{\degree}$. 
It is consistent with the current world average with a better precision. 
\end{abstract}
\maketitle

\section{Introduction}

The test of the standard model explanation of CP violation, which is accommodated by a complex phase in the Cabibbo-Kobayashi-Maskawa (CKM) matrix, is the central goal of heavy flavor physics program. Specifically, 
using $B$ decays to determine the three angles $\alpha$, $\beta$ and $\gamma$ of the usual non-squashed unitarity triangle of the CKM matrix respectively and thus to test the closure of the unitarity triangle is a very straightforward and promising way to accomplish this goal. Any discrepancies would suggest possible new sources 
of {\it CP} violation beyond the standard model.

In principle, $\alpha$, $\beta$ and $\gamma$ can be determined via measurements of {\it CP} violating asymmetry in neutral $B$ decays to {\it CP} eigenstates. If a single CKM amplitude or different amplitudes with the same CKM phases contribute to the decay of $B^0$ meson, then the mixing-induced {\it CP} asymmetry is a pure function of CKM parameters, which are both from the two neutral $B$ mesons mixing and the $B^0$ decay, with no strong phase uncertainties.  
As it is well  known  that the angle $\beta$ can be determined in a reliable way with the help of the mixing-induced {\it CP} violation of a single "gold-plated" mode $B^0 \to J/\psi K_S$. 
Likewise, for $\alpha$, it can be extracted using neutral B decay, $B^0 \to \pi^+ \pi^-$,  using the isospin symmetry analysis to separate the strong phase difference of tree and penguin contributions by other 
$B \to \pi \pi$ decays.
Theoretically, similar with the measurement of $\beta$ and $\alpha$, a straightforward way to obtain of 
$\gamma$ might be to use CKM-suppressed $B_s^0$ decay, $B^0 \to \rho K_S$, or a analysis for the decays 
 $B_s^0 \to D^0 \phi,\bar{D}^0 \phi$ and $D_1^0 \phi$, as in \cite{Gronau:1990ra}.
 However, the observed mixing-induced {\it CP} asymmetries are expected to be strongly diluted by the large 
 $B_s-\bar{B_s}$ mixing, so that to determine $\gamma$ in this way is considerably more involved than 
 $\beta$ and $\alpha$.

The third angle $\gamma$ is currently the least known. It usually depends on strong phase difference of different 
$B$ decays, which is difficult to calculate reliably.  
One of the theoretically cleanest way of determining $\gamma$ is to utilize the interference between the $b \to c \bar{u} s$ and $b \to u\bar{c}s$ decay amplitudes with the intermediate states $D^0$ and $\bar{D^0}$ mesons subsequently decay to common final states rather than to use $B^- \to D^0 K_S$ and $B^- \to \bar{D^0} K_S$ decays directly, due to the large uncertainties of the two amplitudes ratio $r_B$ and strong phase difference between them. According to different common final states, the methods can be divided into: the GLW \cite{Gronau:1991dp} method, with $D$ meson decaying to CP eigenstate; The ADS \cite{Atwood:1996ci} method, with the final state not CP eigenstate but using doubly Cabibbo-suppressed decays to enhance the CP violation effect; The GGSZ \cite{Giri:2003ty} method, which exploits the three-body $D$ decays to self-conjugate modes, such as $D^0 \to K_S  K^+ (\pi^+) K^-(\pi^-)$.
Since the hadronic parameters involved in these decays are not yet known, it is still not clear which of the proposed methods above is more sensitive to $\gamma$. The world average values are mainly achieved by combining the various methods above with more decay modes involved to decrease the statistical uncertainties, which are  $\gamma = (71.1^{+4.6}_{-5.3})^{\circ}$~\cite{Amhis:2016xyh}, $\gamma = (73.5^{+4.2}_{-5.1})^{\circ}$~\cite{Charles:2015gya} and $\gamma = (70.0 \pm 4.2)^{\circ}$~\cite{Bona:2006ah} fitted by HFLAV, CKMFitter and UTfit Collaboration, respectively. The latest combination of $\gamma$ measurements by the LHCb collaboration yield $(74.0^{+5.0}_{-5.8})^{\degree}$ \cite{LHCbgamma}.
As there is no penguin diagram pollution in these charmed $B$ decays and the source of theoretical uncertainty on $\gamma$ determined from higher-order electroweak corrections is also very small, a shift $\delta \gamma \lesssim \mathcal{O}(10^{-7})$ calculated in \cite{Brod:2013sga}, the uncertainty of approximately $5^{\degree}$ on $\gamma$ are statistically limited. The reason for these relatively large statistical uncertainties is, for instance, Br($B^- \to \bar{D^0} K^-$)  $\sim $ $\mathcal{O}(10^{-6})$ suffers from some serious experimental difficulties. 

 The two-body charmless  $B$ meson decays receive both contributions from tree and penguin diagrams with relatively large branching ratios, at the order of $\mathcal{O}(10^{- 6}-10^{- 5})$. Their branching ratios and CP asymmetry parameters depend strongly on the interference of tree and penguin diagrams with different weak and strong phases. This provides a possible way to measure the CKM $\gamma$.  The only problem here is how to calculate or extract the different strong phases between tree and penguin diagrams of charmless $B$ decays reliably. The methods proposed in ref.\cite{Gronau:1994bn,Gronau:1995hn,Neubert:1998pt} extract the strong phases in $B\to \pi \pi$, $B \to \pi K$ and $B \to  KK$ decays by applying flavor SU(3) symmetry. Fleischer propose method to use the decays $B_d \to \pi^+ \pi^- $ and $B_s \to  K^+ K^-$ through the U-spin flavor symmetry of strong interactions \cite{Fleischer:1999pa}. All these methods require a number of experimental measurements but do not depend on the non-perturbative QCD calculation. However the precision on determination of $\gamma$ is limited by the theoretical uncertainties from the flavor SU(3) breaking effects or U-spin-breaking corrections. Some of the methods can only provide bound on $\gamma$, which serves complementary indirect constraint to the the unitarity triangle.
Recently, three-body charmless $B$ decays, whose amplitudes are also related by flavor SU(3) symmetry 
\cite{ReyLeLorier:2011ww,Bhattacharya:2013cla,Bertholet:2018tmx} or U-spin flavor symmetry
\cite{Bhattacharya:2015uua}, are utilized to extract the CKM angle $\gamma$. 
The uncertainty of this fit is around the order of $10^{\degree}$ with six possible solutions found in the latest paper\cite{Bertholet:2018tmx}. 

In order to improve the  precision  of $\gamma$ angle measurements, one has to deal with the  flavor SU(3) breaking effect.
Recently, the factorization-assisted-topological-amplitude  approach is proposed in \cite{Zhou:2016jkv,Li:2012cfa,Li:2013xsa,Zhou:2015jba} to parameterize all the contributions in charmless B decays by topological diagrams, but keep most of the SU(3) breaking effects. 
Like the previous version of topological diagram approach \cite{Cheng:2014rfa}, most of the hadronic decay amplitudes of the weak diagrams are fitted from the experimental measurements instead of perturbative QCD calculation. Thus, it is  model-independent.
But we also take into account the flavor SU(3) breaking effects in each flavor topological diagram characterized by   different decay constants and weak transition form factors. As a result, we can reduce the number of the unknown hadronic parameters by fitting all the charmless two-body $B \to PP$, $PV$ decays together \cite{Zhou:2016jkv}, while the previous one \cite{Cheng:2014rfa} can only fit the $B \to PP$,  and $B \to PV$ decay modes, separately with two sets of parameters.
In the present work, we will try to use all the experimental measured $B \to PP$, $PV$ decays observables to do the global fit again but leaving the weak phase $\gamma$ to be fitted from the abundant experimental data together with the hadronic parameters. 
There are also a number of measured B decay channels with two vector meson final states. Since this kind of decays are more complicated with transverse polarization degrees, we will not include them in our current study to introduce more free parameters.
Explicitly, in the present work, we will fit 15 parameters from 37 experimental measured branching fractions and 11 {\it CP} asymmetry parameters of $B \to PP$, $PV$ decays. 
 
We begin in Sec.\ref{FAT} with a summary of the parameterization of tree and penguin topological amplitudes of charmless $B \to PP$, $PV$ decays, leaving one of the weak phases, $\gamma$, as a free parameter to be fitted with hadronic parameters together. 
The fitted result of CKM angle $\gamma$ with experimental and theoretical uncertainties is presented in Sec.\ref{results}. Sec.\ref{conclusion} is the conclusion.

\section{Parameterization of Decay Amplitudes for different Topological diagrams}\label{FAT}

\begin{figure}[htbp]
\begin{center}
\vspace{-3cm}
\scalebox{0.9}{\epsfig{file=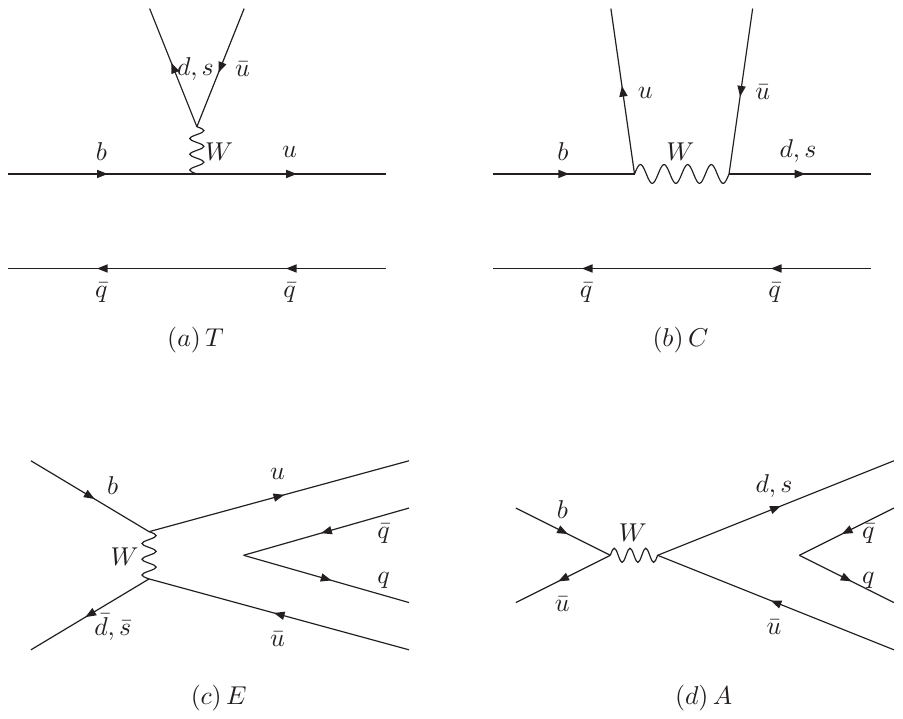}}
\vspace{-17cm}
\caption{Topological tree diagrams contributing to
         $B\to PP $ and $B\to PV$ decays:
 (a) the color-favored tree emission diagram $T$,
 (b) the color-suppressed tree emission diagram  $C$,
 (c) the $W$-exchange diagram $E$ and
 (d) the $W$-annihilation diagram $A$.}
\label{Tree}
\end{center}
\end{figure}

\begin{figure}[htbp]
\begin{center}
\vspace{-4cm}
\scalebox{0.9}{\epsfig{file=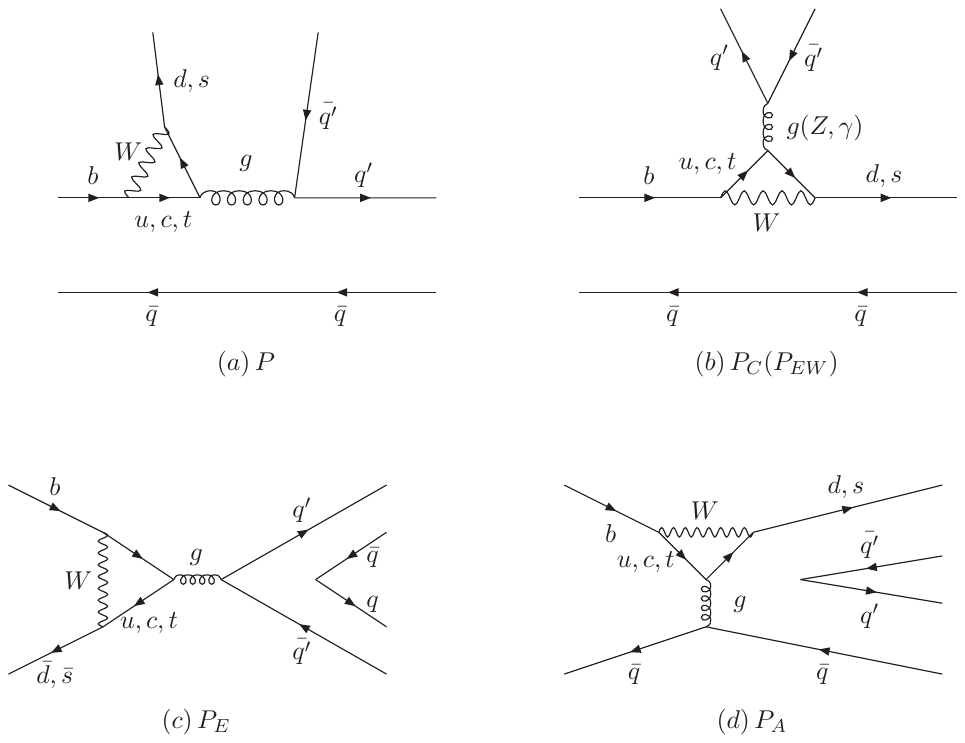}}
\vspace{-16cm}
\caption{Topological penguin diagrams contributing to
              $B\to PP $ and $B\to PV$ decays:
  (a) the  QCD-penguin emission diagram $P$,
  (b) the flavor-singlet   QCD-penguin diagram $P_C$
        or EW-penguin   diagram $P_{EW}$,
  (c) the time-like   QCD-penguin diagram $P_E$ and
  (d) the  space-like  QCD-penguin annihilation diagram $P_A$.}
\label{Penguin}
\end{center}
\end{figure}

The charmless two body $B$ meson decays are induced by weak interactions through tree diagram and penguin diagram  in the quark level. For different $B$ decay final states, the tree level weak decay diagram can contribute via: the so-called  color-favored tree emission diagram $T$,
color-suppressed tree emission diagram $C$,
$W$-exchange tree diagrams $E$ and
$W$ annihilation tree diagrams $A$, which are shown in Fig.\ref{Tree}, respectively. 
The 1-loop corrections from QCD penguin diagrams are not really suppressed due to a larger CKM matrix element comparing with the tree diagram. They are also grouped into four categories:
  (a) the  QCD-penguin emission diagram $P$,
  (b) the flavor-singlet   QCD-penguin diagram $P_C$
  or EW-penguin   diagram $P_{EW}$,
  (c) the time-like   QCD-penguin diagram $P_E$ and
  (d) the  space-like  QCD-penguin annihilation diagram $P_A$, shown in Fig.\ref{Penguin}.
  
The color-favored tree ($T$) topology shown in Fig.\ref{Tree}(a) is proved factorization to all orders of 
$\alpha_{s}$ expansion in QCD factorization approach \cite{Beneke:2000ry}, the perturbative QCD factorization approach \cite{Lu:2000em} and the soft-collinear effective theory \cite{Bauer:2000yr}, thus its formula is easily parametrized as
\begin{align}\label{eq:T}
T^{P_{1}P_{2}}&=i\frac{G_{F}}{\sqrt{2}}V_{ub}V^*_{uq^{'}}a_{1}
(\mu)f_{p_{2}}(m_{B}^{2}-m_{p_{1}}^{2})F_{0}^{BP_{1}}(m_{p_{2}}^{2}),\nonumber\\
T^{PV}&=\sqrt{2}G_{F}V_{ub}V^*_{uq^{'}}a_{1}
(\mu)f_{V} m_{V}F_{1}^{B-P}(m_{V}^{2})(\varepsilon^{*}_{V}\cdot p_{B}),\nonumber\\
T^{VP}&=\sqrt{2}G_{F}V_{ub}V^*_{uq^{'}}a_{1}
(\mu)f_{P} m_{V}A_{0}^{B-V}(m_{P}^{2})(\varepsilon^{*}_{V}\cdot p_{B}),
\end{align}
where the superscript  of $T^{P_{1}P_{2}}$ denote the final mesons
are two pseudoscalar mesons, $T^{PV(VP)}$ for recoiling mesons are
pseudoscalar meson (vector meson). $q' =d,s$ quark for $\Delta S=0,1$ transition, respectively.
$a_1(\mu)$ is the effective Wilson coefficient from short distance QCD corrections, 
$a_1(\mu)=C_2(\mu)+C_1(\mu)/3 =1.05$ at next-to-leading order \cite{Li:2005kt},
with factorization scale $\mu$, insensitive to different final state mesons, 
chosen at $\mathrm{m_b}/2=2.1 \mathrm{GeV}$ usually. 
The decay constants $f_{P}$, $f_{V}$ and form factors $F_{0}^{BP_{1}}$, $F_{1}^{B-P}$ and $A_{0}^{B-V}$ characterize the $SU(3)$ breaking effects. $\varepsilon^{*}_{V}$ is the polarization vector of vector meson and $p_{B}$ is the 4-momentum of $B$ meson.

For the   non-factorization dominant color suppressed tree diagrams, we parameterize them as
\begin{align}\label{eq:C}
C^{P_{1}P_{2}}&=i\frac{G_{F}}{\sqrt{2}}V_{ub}V^*_{uq^{'}}\chi^{C}\mathrm{e}^{i\phi^{C}}
        f_{p_{2}}(m_{B}^{2}-m_{p_{1}}^{2})F_{0}^{BP_{1}}(m_{p_{2}}^{2}),\nonumber\\
C^{PV}&=\sqrt{2}G_{F}V_{ub}V^*_{uq^{'}}\chi^{C^{\prime}}\mathrm{e}^{i\phi^{C^{\prime}}}
         f_{V} m_{V}F_{1}^{B-P}(m_{V}^{2})(\varepsilon^{*}_{V}\cdot p_{B}),
\nonumber\\
C^{VP}&=\sqrt{2}G_{F}V_{ub}V^*_{uq^{'}}\chi^{C}\mathrm{e}^{i\phi^{C}}
        f_{P} m_{V}A_{0}^{B-V}(m_{P}^{2})(\varepsilon^{*}_{V}\cdot p_{B}),
\end{align}
 where $\chi^{C}$, $\phi^{C}$ represent the magnitude and associate phase 
 of pseudo-scalar meson emitted decays $B\to PP$, $VP$. 
The prime in $\chi^{C^{\prime}}$, $\phi^{C^{\prime}}$ denote differences with respect to 
the hadronic parameter in the amplitude of vector meson emitted decays $B\to PV$.
Similarly, we parameterize    the W-exchange diagrams as 
\begin{align}\label{eq:E}
E^{P_{1}P_{2}} &=i\frac{G_{F}}{\sqrt{2}}V_{ub}V^*_{uq^{'}} \chi^{E} \mathrm{e}^{i\phi^{E}}
f_{B}m_{B}^{2}(\frac{f_{p_{1}}f_{p_{2}}}{f_{\pi}^{2}}),\nonumber\\
E^{PV,VP} &=\sqrt{2}G_{F}V_{ub}V^*_{uq^{'}}\chi^{E} \mathrm{e}^{i\phi^{E}}
f_{B}m_{V}(\frac{f_{P}f_{V}}{f_{\pi}^{2}})(\varepsilon^{*}_{V}\cdot p_{B}),
\end{align}
with $\chi^{E}$, $\phi^{E}$ to represent the magnitude and   strong phase. 
We will ignore the W annihilation topology, as its contribution is negligible as discussed in \cite{Cheng:2014rfa}.
 
The penguin emission diagram is also proved factorization in various QCD-inspired approaches and the soft-collinear effective theory to all orders in $\alpha_S$ expansion. Thus its amplitudes are as following:
\begin{align}\label{eq:P}
P^{PP}&=-i\frac{G_{F}}{\sqrt{2}}V_{tb}V_{tq^{'}}^{*}\left[a_{4}(\mu)+\chi^{P}\mathrm{e}^{i\phi^{P}}r_{\chi}\right]
f_{p_{2}}(m_{B}^{2}-m_{p_{1}}^{2})F_{0}^{BP_{1}}(m_{p_{2}}^{2}),\nonumber
\\
P^{PV}&=-\sqrt{2}G_{F} V_{tb}V_{tq^{'}}^{*}a_{4}(\mu) f_{V}m_{V}F_{1}^{B-P}m_{V}^{2}
(\varepsilon^{*}_{V}\cdot p_{B}),\nonumber\\
P^{VP}&=-\sqrt{2}G_{F}V_{tb}V_{tq^{'}}^{*}\left[a_{4}(\mu)-\chi^{P} \mathrm{e}^{i\phi^{P}}r_{\chi}\right]
 f_{P}m_{V}A_{0}^{B-V}(m_{P}^{2})(\varepsilon^{*}_{V}\cdot p_{B}),
\end{align}
where $\chi^P$ and $\phi^P$  denote the ``chiral enhanced" penguin contributions,  
with the chiral factor of pseudo-scalar meson $r_{\chi}$. The Wilson coefficient $a_{4}(\mu)$ of effective penguin operators are calculated to the next-to-leading order  \cite{Li:2005kt}.
We parameterize the flavor-singlet penguin diagram as
\begin{align}\label{eq:PC}
P_{C}^{PP}&=-i\frac{G_{F}}{\sqrt{2}}V_{tb}V_{tq^{'}}^{*}\chi^{P_C}\mathrm{e}^{i\phi^{P_C}}
 f_{p_{2}}(m_{B}^{2}-m_{p_{1}}^{2})F_{0}^{BP_{1}}(m_{p_{2}}^{2}) ,\nonumber\\
P_{C}^{PV}&=-\sqrt{2}G_{F}V_{tb}V_{tq^{'}}^{*} \chi^{P_C^{\prime}}\mathrm{e}^{i\phi^{P_C^{\prime}}}
f_{V}m_{V}F_{1}^{B-P}(m_{V}^{2})(\varepsilon^{*}_{V}\cdot p_{B}),\nonumber\\
P_{C}^{VP}&=-\sqrt{2}G_{F}V_{tb}V_{tq^{'}}^{*} \chi^{P_C}\mathrm{e}^{i\phi^{P_C}}
f_{P}m_{V}A_{0}^{B-V}(m_{P}^{2})(\varepsilon^{*}_{V}\cdot p_{B}).
\end{align}

Looking carefully at the  Fig.\ref{Penguin}(a) and \ref{Penguin}(d), the only difference between them is the hard gluon attached to different light quark pair. Since we do not calculate the QCD corrections, but fitted from experimental measurements, these two diagrams will give the same contribution. 
Since the contribution of pseudo-scalar meson emission $P_{A}^{PP,VP}$ is already encoded in the parameter $r_{\chi}\chi^{P}$, $\mathrm{e}^{i\phi^{P}}$ in Eq.(\ref{eq:P}) for diagram Fig.\ref{Penguin}(a), we have only one contribution left for space-like penguin diagrams: the vector meson emission one:
\begin{align}\label{eq:PA}
P_{A}^{PV}&=-\sqrt{2}G_{F}V_{tb}V_{tq^{'}}^{*}\chi^{P_{A}}\mathrm{e}^{i\phi^{P_{A}}}
f_{B}m_{V}(\frac{f_{P}f_{V}}{f_{\pi}^{2}})(\varepsilon^{*}_{V}\cdot p_{B}).
\end{align}
The contribution from time-like penguin ($P_E$) diagram is argued small, 
which can be ignored   in decay modes not dominated by it \cite{Zhou:2016jkv}.
 
The electroweak penguin topology ($P_{EW}$) is similar to the tree and penguin emission diagrams, 
which can be proved factorization. We calculate this diagram in QCD as
\begin{align}\label{eq:PEW}
P_{EW}^{PP}&=-i\frac{G_{F}}{\sqrt{2}}V_{tb}V_{tq^{'}}^{*}e_{q}\frac{3}{2}a_{9}(\mu)
f_{p_{2}}(m_{B}^{2}-m_{p_{1}}^{2})F_{0}^{BP_{1}}(m_{p_{2}}^{2}),\nonumber\\
P_{EW}^{PV}&=-\sqrt{2}G_{F}V_{tb}V_{tq^{'}}^{*}e_{q}\frac{3}{2}a_{9}(\mu)
f_{V}m_{V}F_{1}^{B-P}(m_{V}^{2})(\varepsilon^{*}_{V}\cdot p_{B}),\nonumber\\
P_{EW}^{VP}&=-\sqrt{2}G_{F}V_{tb}V_{tq^{'}}^{*}e_{q}\frac{3}{2}a_{9}(\mu)
f_{P}m_{V}A_{0}^{B-V}(m_{P}^{2})(\varepsilon^{*}_{V}\cdot p_{B}),
\end{align}
where $a_9(\mu)$ is the effective Wilson coefficient at the next-to-leading order accuracy.


\section{Numerical Results}\label{results}

From the equations (\ref{eq:T}-\ref{eq:PEW}) in the previous section, one notices that all the tree amplitudes are proportional to $ V_{ub}V^*_{uq^{'}}$; while the penguin amplitudes are proportional to $V_{tb}V_{tq^{'}}^{*} = - (V_{ub}V_{uq^{'}}^{*}+V_{cb}V_{cq^{'}}^{*})$.  Except $V_{ub} \equiv |V_{ub}|\, \mathrm{e}^{-i \gamma}$, 
all other CKM matrix elements $V_{uq^{'}}^{*}$, $V_{cb}V_{cq^{'}}^{*}$ are approximately real numbers without electroweak phase. The phase angle $\gamma$ is a free parameter to be fitted from experimental data.
The magnitudes of {CKM} matrix elements  are input parameters cited from ref.\cite{Tanabashi:2018oca}
\begin{align}
|V_{ud}| &=0.97420\pm 0.00021\, ,~~~|V_{us}| =0.2243\pm 0.0005\, ,~~~|V_{ub}| =0.00394\pm 0.00036\,, \nonumber \\
|V_{cd}| &=0.218\pm 0.004\, ,~~~~~~~~~|V_{cs}| =0.997\pm 0.017\, ,~~~~~~~|V_{cb}|=0.0422\pm 0.0008\, .
\end{align}
The remaining parameters expressed in decay amplitude formulas are the meson decay constants and transition form factors.
The meson decay constants are measured by experiments or calculated by theoretical approaches, such as 
covariant light front approach \cite{Cheng:2003sm}
light-cone sum rules \cite{Ball:2006eu,Straub:2015ica},
QCD sum rules \cite{Gelhausen:2013wia,Penin:2001ux} etc. 
We show the values in Table \ref{tab:decay constants}
mostly in average by PDG \cite{Tanabashi:2018oca}.  
 
\begin{table}
\caption{The decay constants of light pseudo-scalar mesons  and vector mesons  (in unit of MeV).}
\label{tab:decay constants}
\begin{center}
\begin{tabular}{ccccccccc}
\hline\hline
$f_{\pi}$ & $f_{K}$  & $f_{B}$ & $f_{\rho}$ & $f_{K^{*}}$ & $f_{\omega}$ &$f_{\phi}$&
\\\hline
$130.2 \pm 1.7$~~& $155.6 \pm 0.4$~~ &$190.9 \pm 4.1$ ~~&  $213 \pm 11$~~ & $220 \pm 11$~~& $192\pm 10$~~ &$225 \pm 11$&
\\
\hline\hline
\end{tabular}
\end{center}
\end{table}

The transition form factors of $B$ meson decays are usually measured through semileptonic B decays modes together with   CKM matrix elements.  Theoretically,  they are calculated in various   approaches: the constitute quark model and light cone quark model \cite{Melikhov:2000yu,Geng:2001de,Lu:2007sg,Albertus:2014bfa},
Covariant light front approach(LFQM) \cite{Cheng:2003sm,Cheng:2009ms,Chen:2009qk},
light-cone sum rules \cite{Ball:1998kk,Ball:2001fp,Ball:1998tj,Ball:2004ye,Ball:2004rg,Khodjamirian:2006st,Bharucha:2010im,Bharucha:2012wy,
Ball:2007hb,Charles:1998dr,Wu:2006rd,Duplancic:2008ix,Meissner:2013hya,Wang:2015vgv,Wu:2009kq,Khodjamirian:2011ub,Ivanov:2011aa,
Ahmady:2014sva,Fu:2014uea,Straub:2015ica,Shen:2016hyv,Lu:2018cfc,Shen:2019zvh,Gao:2019lta},
 PQCD \cite{Li:2012nk,Wang:2012ab,Wang:2013ix,Fan:2013qz,
Fan:2013kqa,Kurimoto:2001zj,Lu:2002ny,Wei:2002iu,Huang:2004hw,Shen:2019vdc}
 and lattice QCD \cite{Horgan:2013hoa,Dalgic:2006dt,Aoki:2013ldr} etc. 
 We combine these results and use the average of the transition form factors of $B$ meson decays at $q^{2}$=0, shown in Table~\ref{tab:formfactor}.
The $q^{2}$ dependence of the transition form factors of $B$ meson decays are described 
with the dipole parametrization,
\begin{equation}\label{eq:ffdipole}
F_{i}(q^{2})={F_{i}(0)\over 1-\alpha_{1}{q^{2}\over M_{\rm pole}^{2}}+
\alpha_{2}{q^{4}\over M_{\rm pole}^{4}}},
\end{equation}
where $F_{i}$ denotes form factors $F_{0}$, $F_{1}$, or $A_{0}$; while $M_{\rm pole}$ is 
the mass of the corresponding pole state,
such as $B_{(s)}$ for $A_{0}$, and $B^{*}_{(s)}$ for $F_{0,1}$. The $q^2$ of charmless B meson decays is not far away from zero, thus the uncertainty of dipole model parameters is neglected in our calculation.
These dipole model parameters are also listed in Table~\ref{tab:formfactor}.

\begin{table}
\caption{The transition form factors of $B$ meson decays at $q^{2}$=0
and dipole model parameters. }\label{tab:formfactor}
\newsavebox{\tablebox}
\begin{lrbox}{\tablebox}
\centering
\begin{tabular}{|c||c|c|c||c|c|c||c|c|c|}
\hline

\hline
           &$~~~F_{0}^{B \to\pi}~~~$ &$~~~F_{0}^{B\to K}~~~$ 
          & $~~~F_{0}^{B \to\eta_{q}}~~~$ &$~~~F_{1}^{B \to\pi}~~~$ &$~~~F_{1}^{B\to K}~~~$ 
           & $~~~F_{1}^{B \to\eta_{q}}~~~$&$~~~A_{0}^{B \to\rho}~~~$ &$~~~A_{0}^{B\to \omega}~~~$ &
           $~~~A_{0}^{B\to K^{*}}~~~$ \\

\hline
$F_i(0)$     & $0.28 \pm 0.03$ & $0.31\pm 0.03$  &  $0.21 \pm 0.02$   & $0.28\pm 0.03$ & $0.31\pm 0.03$  & $0.21 \pm 0.02$   &$ 0.36  \pm 0.04$ & $0.32 \pm 0.03$ & $0.39\pm 0.04 $    \\
$\alpha_1$ & 0.50 & 0.53   & 0.52 & 0.52 & 0.54 &  1.43   & 1.56 & 1.60  & 1.51   \\
$\alpha_2$ & -0.13 & -0.13  &   0  & 0.45 & 0.50&  0.41  & 0.17 & 0.22 &0.14   \\
\hline
\end{tabular}
\end{lrbox}
 \scalebox{0.84}{\usebox{\tablebox}}
\end{table}

To minimize the statistical uncertainties, we should use the maximum amount of experimental observables of 
$B \to PP$, $PV$ decays. However, some of them are measured with very poor precision,  therefore 
we will not use those measurements with less than $3\sigma$ significance  in the following fitting program. 
Then we have 37 branching ratios and 11 {\it CP} violation observations of $B \to PP$, $PV$ decays 
from the current experimental data in ref.~\cite{Tanabashi:2018oca} and 2019 update in the website.

We use the $\chi^2$ fit method by Miniut program~\cite{James:2004}, where the $\chi^2$ function in terms of n experimental observables $x_i \pm \Delta x_i$ and the corresponding theoretical results $x_i^{\mathrm{th}}$ defined as
\begin{align}
\chi^{2}=\sum_{i=1}^{n}\left(\frac{x_i^{\rm th}-x_i}{\Delta x_i}\right)^2 .
\end{align}
The corresponding theoretical results are written as functions of those 15 unknown theoretical parameters in topological amplitudes. The best-fitted parameters are
\begin{align}\label{parameter}
\gamma=(69.8 \pm 2.1)^{\degree} & \nonumber \\
\chi^{C}=0.41 \pm 0.06,~~~&\phi^{C}=-1.74 \pm 0.09,\nonumber \\
\chi^{C^{\prime}}=0.40 \pm 0.17,~~~&\phi^{C^{\prime}}=1.78\pm 0.10,\nonumber \\
\chi^{E}=0.06\pm0.006,~~~&\phi^{E}=2.76\pm 0.13,\nonumber\\
\chi^{P}=0.09\pm0.003,~~~&\phi^{P}=2.55\pm 0.03\nonumber \\
\chi^{P_C}=0.045 \pm 0.003,~~~&\phi^{P_C}=1.53 \pm 0.08,\nonumber \\
\chi^{P_C^{\prime}}=0.037\pm 0.003,~~~&\phi^{P_C^{\prime}}=0.67 \pm 0.08,\nonumber \\
\chi^{P_A}=0.006\pm0.0008,~~~&\phi^{P_A}=1.49\pm 0.09,
\end{align}
with $\chi^{2}/\text{d.o.f}=45.4/33=1.4$. 
The uncertainties  shown here for all the parameters are produced from the $\chi^2$ fit program Miniut, 
which are mainly transmitted from the experimental statistical and systematic uncertainty.  
There should also be theoretical uncertainties on $\gamma$ extraction.
The major source of theoretical uncertainties in our calculation are
the uncertainties of input parameters: $|V_{ub}|$, $|V_{cb}|$, hadronic form factors and decay constants.
We repeat the fit program above but with the input parameters above varying by following a Gaussian distribution one by one. Then these theoretical uncertainties on $\gamma$ can be assessed through the distribution of central values of $\gamma$ due to the variations of these input parameters.
The values of theoretical uncertainty $\sigma_{\mathrm{(T.)}}$ obtained are $ 0.2^{\degree}$,  $ 0.2^{\degree}$, $0.9^{\degree}$ and $ 0.1^{\degree}$ originated from the uncertainties of $|V_{ub}|$,$|V_{cb}|$, form factors and decay constants, respectively. The total theoretical uncertainty is $ 0.9^{\degree}$.
Our final result of $\gamma$ is then $(69.8 \pm 2.1 \pm 0.9) ^{\degree}$, 
which is in good agreement with the current world averages: $\gamma = (71.1^{+4.6}_{-5.3})^{\circ}$~\cite{Amhis:2016xyh}, $\gamma = (73.5^{+4.2}_{-5.1})^{\circ}$~\cite{Charles:2015gya} 
and $\gamma = (70.0 \pm 4.2)^{\circ}$~\cite{Bona:2006ah} and the measurement of 
$(74.0^{+5.0}_{-5.8})^{\degree}$ by the latest LHCb collaboration \cite{LHCbgamma}. 
It is obvious that the uncertainties of $\gamma$ have been shrunk roughly half of the uncertainties 
on the world-average values.

\section{Conclusion}\label{conclusion}

The charmless B meson decays receive contributions from both of the tree amplitudes and the loop penguin amplitudes. The interference between the two amplitudes makes the branching ratios of these decay modes sensitive to the CKM angles, where large direct CP asymmetries are observed. Since non-perturbative dynamics involved, the hadronic matrix elements of these decays are always difficult to calculate precisely. We try to  parametrize the decay amplitudes into different topological diagrams, which can be fitted from the experimental measured quantities, such as branching ratios and CP asymmetry parameters. To improve the precision of the global fit, we factorize the corresponding decay constant and form factors to characterize the flavor SU(3) breaking effect. 
We extract the CKM weak angle $\gamma$ using all the measured two body charmless $B\to PP$, $PV$ decays 
in factorization assisted topological amplitude approach. 
The determined value is $(69.8 \pm 2.1 \pm 0.9) ^{\degree}$, with
the first uncertainty  translated from the experimental error of decay channels and the second error from hadronic parameter and CKM matrix elements.
The result of $\gamma$ is well compatible with the current world average value and the measurement of 
$(74.0^{+5.0}_{-5.8})^{\degree}$ by the latest LHCb collaboration, but with less uncertainty.

\section*{Acknowledgments}
We are grateful to Wen-Bin Qian and Xiao-Rui Lyu for useful discussion.
The work is partly supported by National Science Foundation of China (11847040, 11521505 and 11621131001) and Scientific Research Foundation of Inner Mongolia University.


\begin{thebibliography}{99}
\bibitem{Gronau:1990ra} 
  M.~Gronau and D.~London,
  Phys.\ Lett.\ B {\bf 253}, 483 (1991).

\bibitem{Gronau:1991dp} 
  M.~Gronau and D.~Wyler,
  Phys.\ Lett.\ B {\bf 265}, 172 (1991).

\bibitem{Atwood:1996ci} 
  D.~Atwood, I.~Dunietz and A.~Soni,
  Phys.\ Rev.\ Lett.\  {\bf 78}, 3257 (1997)
  [hep-ph/9612433].
  
\bibitem{Giri:2003ty} 
  A.~Giri, Y.~Grossman, A.~Soffer and J.~Zupan,
  Phys.\ Rev.\ D {\bf 68}, 054018 (2003)
  [hep-ph/0303187].
\bibitem{Amhis:2016xyh} 
  Y.~Amhis {\it et al.} [HFLAV Collaboration],
  Eur.\ Phys.\ J.\ C {\bf 77}, no. 12, 895 (2017)
  [arXiv:1612.07233 [hep-ex]].
  
\bibitem{Charles:2015gya} 
  J.~Charles {\it et al.},
  Phys.\ Rev.\ D {\bf 91}, no. 7, 073007 (2015)
  [arXiv:1501.05013 [hep-ph]].
  
\bibitem{Bona:2006ah} 
  M.~Bona {\it et al.} [UTfit Collaboration],
  JHEP {\bf 0610}, 081 (2006)
  [hep-ph/0606167].  
  
 \bibitem{LHCbgamma}
   LHCb collaboration, Update of the LHCb combination of the CKM angle using $B \to DK$ decays,
    LHCb-CONF-2018-002.  
  
\bibitem{Brod:2013sga} 
  J.~Brod and J.~Zupan,
  JHEP {\bf 1401}, 051 (2014)
  [arXiv:1308.5663 [hep-ph]].
  
\bibitem{Gronau:1994bn} 
  M.~Gronau, J.~L.~Rosner and D.~London,
  Phys.\ Rev.\ Lett.\  {\bf 73}, 21 (1994)
  [hep-ph/9404282].
  
\bibitem{Gronau:1995hn} 
  M.~Gronau, O.~F.~Hernandez, D.~London and J.~L.~Rosner,
  Phys.\ Rev.\ D {\bf 52}, 6374 (1995)
  [hep-ph/9504327].
  
\bibitem{Neubert:1998pt} 
  M.~Neubert and J.~L.~Rosner,
  Phys.\ Lett.\ B {\bf 441}, 403 (1998)
  [hep-ph/9808493].
  
\bibitem{Fleischer:1999pa} 
  R.~Fleischer,
  Phys.\ Lett.\ B {\bf 459}, 306 (1999)
  [hep-ph/9903456].
  
\bibitem{ReyLeLorier:2011ww} 
 N.~Rey-Le Lorier and D.~London,
  ``Measuring gamma with $B \to K \pi \pi$ and $B \to K K \bar{k}$ Decays,''
  Phys.\ Rev.\ D {\bf 85}, 016010 (2012)
  [arXiv:1109.0881 [hep-ph]].
  
\bibitem{Bhattacharya:2013cla} 
  B.~Bhattacharya, M.~Imbeault and D.~London,
  ``Extraction of the CP-violating phase $\gamma$ using $B \to K \pi \pi$ and $B \to K K {\bar K}$ decays,''
  Phys.\ Lett.\ B {\bf 728}, 206 (2014)
  [arXiv:1303.0846 [hep-ph]].
  
\bibitem{Bertholet:2018tmx} 
  E.~Bertholet, E.~Ben-Haim, B.~Bhattacharya, M.~Charles and D.~London,
  ``Extraction of the CKM phase $\gamma$ using charmless 3-body decays of $B$ mesons,''
Phys. Rev. D 99, 114011 (2019)  [arXiv:1812.06194 [hep-ph]].
  
\bibitem{Bhattacharya:2015uua} 
  B.~Bhattacharya and D.~London,
  ``Using U spin to extract $\gamma$ from charmless $B \to PPP$ decays,''
  JHEP {\bf 1504}, 154 (2015)
  [arXiv:1503.00737 [hep-ph]].
  
\bibitem{Zhou:2016jkv} 
  S.~H.~Zhou, Q.~A.~Zhang, W.~R.~Lyu and C.~D.~Lu,
  Eur.\ Phys.\ J.\ C {\bf 77}, no. 2, 125 (2017)
  [arXiv:1608.02819 [hep-ph]].
  
  \bibitem{Li:2012cfa}
  H.~n.~Li, C.~D.~Lu and F.~S.~Yu,
  Phys.\ Rev.\ D {\bf 86}, 036012 (2012)
  [arXiv:1203.3120 [hep-ph]].
  
\bibitem{Li:2013xsa}
  H.~n.~Li, C.~D.~Lu, Q.~Qin and F.~S.~Yu,
  Phys.\ Rev.\ D {\bf 89}, no. 5, 054006 (2014)
  [arXiv:1305.7021 [hep-ph]].

\bibitem{Zhou:2015jba}
  S.~H.~Zhou, Y.~B.~Wei, Q.~Qin, Y.~Li, F.~S.~Yu and C.~D.~Lu,
  Phys.\ Rev.\ D {\bf 92}, no. 9, 094016 (2015)
  [arXiv:1509.04060 [hep-ph]].
    
  \bibitem{Cheng:2014rfa}
  H.~Y.~Cheng, C.~W.~Chiang and A.~L.~Kuo,
  Phys.\ Rev.\ D {\bf 91}, no. 1, 014011 (2015)
  [arXiv:1409.5026 [hep-ph]].

\bibitem{Beneke:2000ry}
  M.~Beneke, G.~Buchalla, M.~Neubert and C.~T.~Sachrajda,
  Nucl.\ Phys.\ B {\bf 591}, 313 (2000)
  [hep-ph/0006124].

\bibitem{Lu:2000em}
  C.~D.~Lu, K.~Ukai and M.~Z.~Yang,
  Phys.\ Rev.\ D {\bf 63}, 074009 (2001)
  [hep-ph/0004213];
  Y.~Y.~Keum, H.~N.~Li and A.~I.~Sanda,
  Phys.\ Rev.\ D {\bf 63}, 054008 (2001)
  [hep-ph/0004173].
\bibitem{Bauer:2000yr}
  C.~W.~Bauer, S.~Fleming, D.~Pirjol and I.~W.~Stewart,
  Phys.\ Rev.\ D {\bf 63}, 114020 (2001)
  [hep-ph/0011336].

\bibitem{Li:2005kt}
  H.~n.~Li, S.~Mishima and A.~I.~Sanda,
  Phys.\ Rev.\ D {\bf 72}, 114005 (2005)
  [hep-ph/0508041].

\bibitem{Tanabashi:2018oca} 
  M.~Tanabashi {\it et al.} [Particle Data Group],
  Phys.\ Rev.\ D {\bf 98}, no. 3, 030001 (2018).
    

\bibitem{Cheng:2003sm}
  H.~Y.~Cheng, C.~K.~Chua and C.~W.~Hwang,
  Phys.\ Rev.\ D {\bf 69}, 074025 (2004)
  [hep-ph/0310359].
  
\bibitem{Ball:2006eu}
  P.~Ball, G.~W.~Jones and R.~Zwicky,
  Phys.\ Rev.\ D {\bf 75}, 054004 (2007)
  [hep-ph/0612081].
  
\bibitem{Straub:2015ica}
  A.~Bharucha, D.~M.~Straub and R.~Zwicky,
  arXiv:1503.05534 [hep-ph].


\bibitem{Gelhausen:2013wia}
  P.~Gelhausen, A.~Khodjamirian, A.~A.~Pivovarov and D.~Rosenthal,
  Phys.\ Rev.\ D {\bf 88}, 014015 (2013)
  Erratum: [Phys.\ Rev.\ D {\bf 89}, 099901 (2014)]
  Erratum: [Phys.\ Rev.\ D {\bf 91}, 099901 (2015)]
  [arXiv:1305.5432 [hep-ph]].

\bibitem{Penin:2001ux}
  A.~A.~Penin and M.~Steinhauser,
  Phys.\ Rev.\ D {\bf 65}, 054006 (2002)
  [hep-ph/0108110].

\bibitem{Melikhov:2000yu}
  D.~Melikhov and B.~Stech,
  Phys.\ Rev.\ D {\bf 62}, 014006 (2000)
  [hep-ph/0001113].

\bibitem{Geng:2001de}
  C.~Q.~Geng, C.~W.~Hwang, C.~C.~Lih and W.~M.~Zhang,
  Phys.\ Rev.\ D {\bf 64}, 114024 (2001)
  [hep-ph/0107012].

\bibitem{Lu:2007sg}
  C.~D.~Lu, W.~Wang and Z.~T.~Wei,
  Phys.\ Rev.\ D {\bf 76}, 014013 (2007)
  [hep-ph/0701265 [HEP-PH]].

\bibitem{Albertus:2014bfa}
  C.~Albertus,
  Phys.\ Rev.\ D {\bf 89}, no. 6, 065042 (2014)
  [arXiv:1401.1791 [hep-ph]].

\bibitem{Cheng:2009ms}
  H.~Y.~Cheng and C.~K.~Chua,
  Phys.\ Rev.\ D {\bf 81}, 114006 (2010)
  Erratum: [Phys.\ Rev.\ D {\bf 82}, 059904 (2010)]
  [arXiv:0909.4627 [hep-ph]].

\bibitem{Chen:2009qk}
  C.~H.~Chen, Y.~L.~Shen and W.~Wang,
  Phys.\ Lett.\ B {\bf 686}, 118 (2010)
  [arXiv:0911.2875 [hep-ph]].

\bibitem{Ball:1998kk}
  P.~Ball and V.~M.~Braun,
  Phys.\ Rev.\ D {\bf 58}, 094016 (1998)
  [hep-ph/9805422].

\bibitem{Ball:1998tj}
  P.~Ball,
  JHEP {\bf 9809}, 005 (1998)
  [hep-ph/9802394].

\bibitem{Ball:2001fp}
  P.~Ball and R.~Zwicky,
  JHEP {\bf 0110}, 019 (2001)
  [hep-ph/0110115].

\bibitem{Ball:2004ye}
  P.~Ball and R.~Zwicky,
  Phys.\ Rev.\ D {\bf 71}, 014015 (2005)
  [hep-ph/0406232].

\bibitem{Ball:2004rg}
  P.~Ball and R.~Zwicky,
  Phys.\ Rev.\ D {\bf 71}, 014029 (2005)
  [hep-ph/0412079].

\bibitem{Ball:2007hb}
  P.~Ball and G.~W.~Jones,
  JHEP {\bf 0708}, 025 (2007)
  [arXiv:0706.3628 [hep-ph]].

\bibitem{Charles:1998dr}
  J.~Charles, A.~Le Yaouanc, L.~Oliver, O.~Pene and J.~C.~Raynal,
  Phys.\ Rev.\ D {\bf 60}, 014001 (1999)
  [hep-ph/9812358].


\bibitem{Bharucha:2010im}
  A.~Bharucha, T.~Feldmann and M.~Wick,
  JHEP {\bf 1009}, 090 (2010)
  [arXiv:1004.3249 [hep-ph]].

\bibitem{Bharucha:2012wy}
  A.~Bharucha,
  JHEP {\bf 1205}, 092 (2012)
  [arXiv:1203.1359 [hep-ph]].

\bibitem{Khodjamirian:2006st}
  A.~Khodjamirian, T.~Mannel and N.~Offen,
  Phys.\ Rev.\ D {\bf 75}, 054013 (2007)
  [hep-ph/0611193].

\bibitem{Khodjamirian:2011ub}
  A.~Khodjamirian, T.~Mannel, N.~Offen and Y.-M.~Wang,
  Phys.\ Rev.\ D {\bf 83}, 094031 (2011)
  [arXiv:1103.2655 [hep-ph]].

\bibitem{Wang:2015vgv}
  Y.~M.~Wang and Y.~L.~Shen,
  Nucl.\ Phys.\ B {\bf 898}, 563 (2015)
  [arXiv:1506.00667 [hep-ph]].
  
\bibitem{Shen:2016hyv} 
  Y.~L.~Shen, Y.~B.~Wei and C.~D.~Lu,
  Phys.\ Rev.\ D {\bf 97}, no. 5, 054004 (2018)
  [arXiv:1607.08727 [hep-ph]].  
  
\bibitem{Lu:2018cfc} 
  C.~D.~Lu, Y.~L.~Shen, Y.~M.~Wang and Y.~B.~Wei,
  JHEP {\bf 1901}, 024 (2019)
  [arXiv:1810.00819 [hep-ph]].
  
\bibitem{Shen:2019zvh} 
  Y.~L.~Shen, J.~Gao, C.~D.~Lu and Y.~Miao,
  Phys.\ Rev.\ D {\bf 99}, no. 9, 096013 (2019)
  [arXiv:1901.10259 [hep-ph]].

\bibitem{Gao:2019lta} 
  J.~Gao, C.~D.~Lu, Y.~L.~Shen, Y.~M.~Wang and Y.~B.~Wei,
  arXiv:1907.11092 [hep-ph].

\bibitem{Meissner:2013hya}
  U.~G.~Meissner and W.~Wang,
  Phys.\ Lett.\ B {\bf 730}, 336 (2014)
  [arXiv:1312.3087 [hep-ph]].

\bibitem{Wu:2006rd}
  Y.~L.~Wu, M.~Zhong and Y.~B.~Zuo,
  Int.\ J.\ Mod.\ Phys.\ A {\bf 21}, 6125 (2006)
  [hep-ph/0604007].

\bibitem{Wu:2009kq}
  X.~G.~Wu and T.~Huang,
  Phys.\ Rev.\ D {\bf 79}, 034013 (2009)
  [arXiv:0901.2636 [hep-ph]].

\bibitem{Duplancic:2008ix}
  G.~Duplancic, A.~Khodjamirian, T.~Mannel, B.~Melic and N.~Offen,
  JHEP {\bf 0804}, 014 (2008)
  [arXiv:0801.1796 [hep-ph]].

\bibitem{Ivanov:2011aa}
  M.~A.~Ivanov, J.~G.~Korner, S.~G.~Kovalenko, P.~Santorelli and G.~G.~Saidullaeva,
  Phys.\ Rev.\ D {\bf 85}, 034004 (2012)
  [arXiv:1112.3536 [hep-ph]].

\bibitem{Ahmady:2014sva}
  M.~Ahmady, R.~Campbell, S.~Lord and R.~Sandapen,
  Phys.\ Rev.\ D {\bf 89}, no. 7, 074021 (2014)
  [arXiv:1401.6707 [hep-ph]].

\bibitem{Fu:2014uea}
  H.~B.~Fu, X.~G.~Wu and Y.~Ma,
  J.\ Phys.\ G {\bf 43}, no. 1, 015002 (2016)
  [arXiv:1411.6423 [hep-ph]].
\bibitem{Li:2012nk}
  H.~n.~Li, Y.~L.~Shen and Y.~M.~Wang,
  Phys.\ Rev.\ D {\bf 85}, 074004 (2012)
  [arXiv:1201.5066 [hep-ph]].

\bibitem{Wang:2012ab}
  W.~F.~Wang and Z.~J.~Xiao,
  Phys.\ Rev.\ D {\bf 86}, 114025 (2012)
  [arXiv:1207.0265 [hep-ph]].
  
\bibitem{Shen:2019vdc} 
  Y.~L.~Shen, Z.~T.~Zou and Y.~Li,
  Phys.\ Rev.\ D {\bf 100}, no. 1, 016022 (2019)
  [arXiv:1901.05244 [hep-ph]].

\bibitem{Wang:2013ix}
  W.~F.~Wang, Y.~Y.~Fan, M.~Liu and Z.~J.~Xiao,
  Phys.\ Rev.\ D {\bf 87}, no. 9, 097501 (2013)
  [arXiv:1301.0197].

\bibitem{Fan:2013qz}
  Y.~Y.~Fan, W.~F.~Wang, S.~Cheng and Z.~J.~Xiao,
  Chin.\ Sci.\ Bull.\  {\bf 59}, 125 (2014)
  [arXiv:1301.6246 [hep-ph]].

\bibitem{Fan:2013kqa}
  Y.~Y.~Fan, W.~F.~Wang and Z.~J.~Xiao,
  Phys.\ Rev.\ D {\bf 89}, no. 1, 014030 (2014)
  [arXiv:1311.4965 [hep-ph]].

\bibitem{Kurimoto:2001zj}
  T.~Kurimoto, H.~n.~Li and A.~I.~Sanda,
  Phys.\ Rev.\ D {\bf 65}, 014007 (2002)
  [hep-ph/0105003].

\bibitem{Lu:2002ny}
  C.~D.~Lu and M.~Z.~Yang,
  Eur.\ Phys.\ J.\ C {\bf 28}, 515 (2003)
  [hep-ph/0212373].
\bibitem{Wei:2002iu}
  Z.~T.~Wei and M.~Z.~Yang,
  Nucl.\ Phys.\ B {\bf 642}, 263 (2002)
  [hep-ph/0202018].
\bibitem{Huang:2004hw}
  T.~Huang and X.~G.~Wu,
  Phys.\ Rev.\ D {\bf 71}, 034018 (2005)
  [hep-ph/0412417].
\bibitem{Horgan:2013hoa}
  R.~R.~Horgan, Z.~Liu, S.~Meinel and M.~Wingate,
  Phys.\ Rev.\ D {\bf 89}, no. 9, 094501 (2014)
  [arXiv:1310.3722 [hep-lat]].
\bibitem{Dalgic:2006dt}
  E.~Dalgic, A.~Gray, M.~Wingate, C.~T.~H.~Davies, G.~P.~Lepage and J.~Shigemitsu,
  Phys.\ Rev.\ D {\bf 73}, 074502 (2006)
  Erratum: [Phys.\ Rev.\ D {\bf 75}, 119906 (2007)]
  [hep-lat/0601021].
\bibitem{Aoki:2013ldr}
  S.~Aoki {\it et al.},
  Eur.\ Phys.\ J.\ C {\bf 74}, 2890 (2014)
  [arXiv:1310.8555 [hep-lat]].
    
\bibitem{James:2004}
    F. James and M. Winker.$http:www.cern.ch/minuit.$ CERN, May 2004.

\end{thebibliography}

\end{document}